# Unveiling Magnetic Interactions of Ruthenium Trichloride via Constraining Direction of Orbital moments: Potential Routes to Realize Quantum Spin Liquid

Y. S. Hou[1,2], H. J. Xiang[1,2,*], and X. G. Gong[1,2,*]

[1] Key Laboratory of Computational Physical Sciences (Ministry of Education), State Key Laboratory of Surface Physics, Department of Physics, Fudan University, Shanghai 200433, China

[2] Collaborative Innovation Center of Advanced Microstructures, Nanjing, 210093, China

Email: hxiang@fudan.edu.cn, xggong@fudan.edu.cn

**Abstract**

Recent experiments reveal that the honeycomb ruthenium trichloride $\alpha$-RuCl$_3$ is a prime candidate of the Kitaev quantum spin liquid (QSL). However, there is no theoretical model which can properly describe its experimental dynamical response, due to the lack of a full understanding of its magnetic interactions. Here, we propose a general scheme to calculate the magnetic interactions in systems (e.g., $\alpha$-RuCl$_3$) with non-negligible orbital moments by constraining the directions of orbital moments. With this scheme, we put forward a minimal $J_1$-$K_1$-$\Gamma_1$-$J_3$-$K_3$ model for $\alpha$-RuCl$_3$ and find that: (I) The third nearest neighbor (NN) antiferromagnetic Heisenberg interaction $J_3$ stabilizes the zigzag antiferromagnetic order; (II) The NN symmetric off-diagonal exchange $\Gamma_1$ plays a pivotal role in determining the preferred direction of magnetic moments and generating the spin wave gap. Exact diagonalization study on this model shows that the Kitaev QSL can be realized by suppressing the NN symmetric off-diagonal exchange $\Gamma_1$ and the third NN Heisenberg interaction $J_3$. Thus, we not only propose a powerful general scheme for investigating the intriguing magnetism of $J_{\text{eff}}$=1/2 magnets, but also point out future directions for realizing the Kitaev QSL in the honeycomb ruthenium trichloride $\alpha$-RuCl$_3$.

*Introduction.*−Quantum spin liquid is one of the most exotic and elusive topological states of frustrated magnets showing remarkable collective phenomena, such as emergent gauge fields and fractional particle excitations [1-3]. Of particular interest is the S=1/2 Kitaev model on the honeycomb lattice which has an exactly solvable QSL ground state [4]. It was proposed that such model may be realized by the $J_{\text{eff}}$=1/2 state in the honeycomb iridium oxides, for instance, $Na_2IrO_3$ and $\alpha$-$Li_2IrO_3$ [5-10]. Unfortunately, various experiments show that $Na_2IrO_3$ has a zigzag antiferromagnetic (AFM) order below 15 K [5, 9, 10] while $\alpha$-$Li_2IrO_3$ has an incommensurate counter-rotating magnetic order [11].

Excitingly, some up-to-date experiments [2, 12, 13] indicate that the honeycomb ruthenium trichloride $\alpha$-$RuCl_3$ is closer to the Kitaev QSL than the widely studied iridium oxides. It has been demonstrated that $\alpha$-$RuCl_3$ exhibits frustrated magnetic interactions [12] and can be described by the $J_{\text{eff}}$=1/2 state [13, 14] which is the prerequisite for realizing the Kitaev model. Most importantly, Banerjee *et al*. reported that $\alpha$-$RuCl_3$ is a proximate Kitaev QSL magnet because its dynamical response measurements above interlayer energy scale are explained in terms of deconfinement physics expected for the QSL [2].

However, a proper theoretical model which could well capture its collective magnetic measurements is still lacking. On one hand, the Heisenberg-Kitaev (HK) model [15, 16] cannot account for the experimentally observed magnon gap and the high-energy magnetic mode $M_2$ [2]. On the other hand, the pure AFM Kitaev model cannot reproduce the experimentally observed concave nature of the edge lower mode in (***Q***, E) space [2, 17]. Although experiments showed that $\alpha$-$RuCl_3$ has a zigzag AFM order below 7 K and its magnetic moments collinearly lie in the ***ac*** plane [2, 18-20], it is not clear whether the magnetic moments take a direction of 35° or -35° away from the ***a*** axis [18]. To address these key questions, the most fundamental issue is to unveil what kinds of magnetic interactions are important in $\alpha$-$RuCl_3$. Once this issue is pinned down, one can put forward a proper theoretical model to capture the experimental observations and explore how to tune the magnetic interactions so as to experimentally realize the QSL in $\alpha$-$RuCl_3$.

In this Letter, to unveil the magnetic interactions of $\alpha$-$RuCl_3$, we propose a novel scheme to extract the magnetic interaction parameters by constraining the directions of orbital moments in the density functional theory (DFT) calculations. Calculations show the magnetic ground state is the zigzag AFM order with magnetic moments along the direction of 35° away from the ***a*** axis, in agreement with experiments [2]. Apart from the NN ferromagnetic (FM) Kitaev interaction $K_1$ and AFM Heisenberg interaction $J_1$, the NN symmetric off-diagonal exchange $\Gamma_1$, and the third NN Heisenberg interaction $J_3$ and Kitaev interaction $K_3$ are surprisingly large. In view of such results, we propose a minimal $J_1$-$K_1$-$\Gamma_1$-$J_3$-$K_3$ model. We find that the zigzag AFM order is stabilized by the third NN AFM Heisenberg interaction $J_3$ and the NN symmetric off-diagonal exchange $\Gamma_1$ is of importance to the selection of the preferred direction of magnetic moments and the generation of the spin wave gap. Exact diagonalization study reveals that the Kitaev QSL can be realized by suppressing the NN symmetric off-diagonal exchange $\Gamma_1$ and the third NN Heisenberg interaction $J_3$. We illustrate that our scheme is powerful for studying the magnetism of $J_{\text{eff}}$=1/2 magnets and shed light on realizing the Kitaev QSL in $\alpha$-RuCl$_3$.

*First-principles Computational details.*−Our first-principles calculations are performed based on the DFT+U method. We use the generalized gradient approximation and the projector-augmented wave (PAW) method with an energy cutoff of 450 eV [21-23]. The on-site repulsion U ranges from 2.5 to 3.5 eV and the Hund coupling is $J_h$ = 0.4 eV [13, 24]. Since Ru has a strong spin-orbit coupling (SOC) [2, 13, 24], SOC is included in our calculations. We adopt the low-temperature monoclinic crystal structure with the space group $C2/m$ [18].

*Method of constraining the directions of orbital moments.*−We first demonstrate why the directions of orbital moments of $\alpha$-RuCl$_3$ should be constrained. Recall that the magnetism of $\alpha$-RuCl$_3$ is described by the $J_{\text{eff}}$=1/2 state [2, 13, 25]. In the low spin $d^5$ configuration in the octahedral crystal field, such as Ru$^{3+}$ in $\alpha$-RuCl$_3$, a hole resides in $t_{2g}$ manifold of $d_{xy}$, $d_{yz}$, and $d_{zx}$ orbitals. The single ion SOC Hamiltonian $H_{SOC} = \lambda \mathbf{L} \cdot \mathbf{S}$ entangles the orbital and spin spaces and gives rise to the novel $J_{\text{eff}}$=1/2 state:

$$\left| J_{\text{eff}} = 1/2, m_{J_{\text{eff}}} = \pm 1/2 \right\rangle = \left[ \left| d_{yz}, \pm\sigma \right\rangle \mp i \left| d_{zx}, \pm\sigma \right\rangle \mp \left| d_{xy}, \mp\sigma \right\rangle \right] / \sqrt{3} \qquad (1).$$

In Eq. (1), where $\sigma$ denotes the spin state. One can obtain based on Eq. (1) that the total magnetic moment (1 $\mu_B$) of the $J_{\text{eff}}$=1/2 state is composed of the dominant orbital moment (2/3 $\mu_B$) and the spin moment (1/3 $\mu_B$) [26]. Therefore the former has a much more vital effect on the magnetic interaction between two different $J_{\text{eff}}$=1/2 states. In spite of the fact that DFT calculations with the spin moments constrained along desired directions were carried out [27, 28], constraining the directions of orbital moments has not been achieved so far. Test calculations of $\alpha$-RuCl$_3$ indicate that the directions of spin and orbital moments will seriously deviate from each other if the directions of the orbital moments are not constrained, and such deviation also appears in $Na_2IrO_3$ and $\alpha$-$Li_2IrO_3$ (Sec. I of Supplemental Material). To extract the magnetic interaction parameters, it is necessary to obtain the total energy of magnetic states with given directions of spin and orbital moments. So it is crucial to constrain the direction of orbital moments in systems with non-negligible orbital moments.

Now let us show how the directions of orbital moments can be constrained in the framework of DFT. To this end, we add to the usual DFT total energy a penalty energy $E_{constr}$:

$$E = E_0 + E_{constr} = E_0 + \sum_t \lambda \left[ \mathbf{L}_l^t - \mathbf{L}_l^{t,0}\left(\mathbf{L}_l^{t,0} \cdot \mathbf{L}_l^t\right)\right]^2 \qquad (2).$$

Here, $E_0$ is the usual DFT total energy, $\mathbf{L}_l^{t,0}$ is a unit vector along the desired direction and $\mathbf{L}_l^t$ is the orbital moment of site $t$ and angular momentum $l$ (e.g., $l = 2$ for Ru 4$d$ orbitals). The parameter $\lambda$ in unit of $\mathrm{eV}/\mu_B^2$ is a Lagrange multiplier that controls the penalty energy $E_{constr}$. This scheme is similar to the method for constraining spin moments proposed by Franchini [27] and Dudarev [29]. A detailed deduction of how this scheme is implemented into the PAW method is given in the Sec. II of Supplemental Material. We find that parameter $\lambda$ ranging from 0.1 to 1.0 $\mathrm{eV}/\mu_B^2$ is sufficient to make the penalty energy $E_{constr}$ less than $10^{-5}$ eV and satisfactorily constrains the directions of orbital moments. Parameter $\lambda = 0.2\,\mathrm{eV}/\mu_B^2$ is employed in the present work. Note that the directions of spin moment are also constrained in our DFT calculations.

*First-principles calculation results.*−We first show the magnetic ground state of $\alpha$-RuCl$_3$ is the zigzag AFM order with magnetic moments along the direction of 35° away from the *a* axis. Experiments reveal that $\alpha$-RuCl$_3$ has a layered honeycomb crystal structure and zigzag AFM order with magnetic moments along a direction of 35° or -35° away from the *a* axis [18] (FIG. 1). In the (*x*, *y*, *z*) coordinate expressing the HK Hamiltonian (FIG. 1a), the directions of 35° and -35° away from the *a* axis are along the [22$\bar{1}$] and *z* axes, respectively. Apart from the zigzag AFM order, we examine other three representative magnetic orders, that is, FM (FIG. 2a), Neel AFM (FIG. 2b) and stripe AFM (FIG. 2c). For each magnetic order, six different directions, namely, *x*, *y*, *z*, *a*, *b*, and [22$\bar{1}$] axes, are taken into account. Shown in FIG. 2e are the energy differences between these magnetic orders as a function of U. It is shown that the zigzag AFM order with magnetic moments along the [22$\bar{1}$] axis (denoted as zigzag-22$\bar{1}$, such denotation method is also applied to other magnetic orders) has the lowest total energy for all considered U cases, which indicates that the magnetic ground state is the zigzag AFM order with magnetic moments along the direction of 35° away from the *a* axis and that *our scheme* works in α-RuCl$_3$.

Having established the magnetic ground state of α-RuCl$_3$, we introduce the Hamiltonian employed in our work. We consider a generalized bilinear Hamiltonian widely used in the study of $J_{\text{eff}}$=1/2 magnet and described as [25, 30-34], up to the third NN Ru-Ru bonds,

$$H = \sum_{ij \in \alpha\beta(\gamma)} \left[ J_{ij} \mathbf{S}_i \cdot \mathbf{S}_j + K_{ij} S_i^\gamma S_j^\gamma + \mathbf{D}_{ij} \cdot (\mathbf{S}_i \times \mathbf{S}_j) + \mathbf{S}_i \cdot \boldsymbol{\Gamma}_{ij} \cdot \mathbf{S}_j \right] \quad (3),$$

where the last term is the generalized symmetric off-diagonal exchange [34] in the form of

$$\mathbf{S}_i \cdot \boldsymbol{\Gamma}_{ij} \cdot \mathbf{S}_j = \Gamma_{ij}^x \left( S_i^y S_j^z + S_i^z S_j^y \right) + \Gamma_{ij}^y \left( S_i^z S_j^x + S_i^x S_j^z \right) + \Gamma_{ij}^z \left( S_i^x S_j^y + S_i^y S_j^x \right).$$

In Eq. (3), $i$ and $j$ label Ru sites and $\mathbf{S}_i$ is a $J_{\text{eff}}$=1/2 pseudo-spin operator with components $S_i^\alpha$ (α=x, y, z). Parameters $J_{ij}$, $K_{ij}$ and $\mathbf{D}_{ij}$ are the Heisenberg, Kitaev and Dzyaloshinskii-Moriya (DM) interactions, respectively. Every Ru-Ru bond $ij$ is distinguished by one spin direction γ (FIG. 1b) and other two directions α and β [34].

α-RuCl$_3$ has the dominating NN FM Kitaev interaction and sizable NN symmetric off-diagonal exchange. Shown in the Table *I* are the calculated magnetic interaction parameters by means of the four-states method [35]. Although the magnitudes of magnetic interaction parameters are U-dependent, the NN x-, y- and z-bonds have FM Heisenberg interactions and dominating FM Kitaev interactions. Note that the NN FM Kitaev interaction is consistent with the recent theoretical results obtained by the nonperturbative exact diagonalization methods [33] and by quantum chemistry calculations [36]. In contrast to our results, the NN Kitaev interaction is estimated to be AFM with the perturbation theory in which tight binding parameters is deduced from the *ab initio* calculations [25]. Furthermore, the NN symmetric off-diagonal exchange is sizable and bond-dependent. For example, the NN z-bond has the sizable $\Gamma^z$ component while the other two components $\Gamma^x$ and $\Gamma^y$ are negligible. Unexpectedly, the second NN magnetic interactions are relatively much weak. However, the third NN Ru-Ru bonds have weak AFM Kitaev interactions and sizable AFM Heisenberg interactions comparable to the NN FM Heisenberg interactions, which is different from the Na$_2$IrO$_3$ case where the second and third NN Heisenberg interactions are both estimated to be comparable to the NN Heisenberg interaction [5].

Lastly, the DM interactions are all negligible because of the qausi-inversion symmetry of the $Ru_2Cl_{10}$ cluster.

*Comparison with experimental observations.* –Based on our calculated magnetic interaction parameters, we propose that the minimal model of $\alpha$-RuCl$_3$ is the $J_1$-$K_1$-$\Gamma_1$-$J_3$-$K_3$ one:

$$H = \sum_{\langle ij \rangle \in \alpha\beta(\gamma)} \left[ J_1 S_i \cdot S_j + K_1 S_i^\gamma S_j^\gamma + \Gamma_1 \left( S_i^\alpha S_j^\beta + S_i^\beta S_j^\alpha \right) \right] + \sum_{\langle\langle\langle ij \rangle\rangle\rangle \in \alpha\beta(\gamma)} \left( J_3 S_i \cdot S_j + K_3 S_i^\gamma S_j^\gamma \right) \quad (4).$$

Here the second NN magnetic interactions are left out of consideration as they are weak compared with the NN and third NN magnetic interactions. For simplicity, the NN (third NN) x-, y- and z- bonds are considered to be equivalent, since they have similar magnetic interactions. The magnitudes of parameters $J_1$, $K_1$, $\Gamma_1$, $J_3$ and $K_3$, listed in the last row in the Table *I*, are obtained by averaging the corresponding results calculated based on the U=3.5 eV.

The experimentally observed zigzag AFM structure is well reproduced by the $J_1$-$K_1$-$\Gamma_1$-$J_3$-$K_3$ model. The efficient exchange Monte Carlo (MC) [37-39] simulation indicates the magnetic ground state is the zigzag AFM order with magnetic moments along the [77$\bar{2}$] direction which is 43$^o$ (close to the experimentally observed 35$^o$) away from the *a* axis. In addition, MC simulations reveal that the magnetic transition temperature is 8.8 K, quite closed to the experimentally measured $T_N$ = 7 K [18].

The third NN AFM Heisenberg interaction and the NN symmetric off-diagonal exchange play an important role in establishing the experimentally observed zigzag AFM structure. If only the dominating NN FM Kitaev interaction $K_1$ and the NN FM Heisenberg interaction $J_1$ are considered, the magnetic ground state is at the boundary of the Kitaev QSL [40]. However, MC simulations show that such proximate Kitaev QSL breaks down and the zigzag AFM order stabilizes provided that the third NN AFM Heisenberg interaction $J_3$ is additionally included. Such simulated result is reasonable because the third NN AFM Heisenberg interaction $J_3$ is magnetically satisfied in the zigzag AFM order (FIG. 1b). Based on such zigzag AFM order, extra MC simulations including the NN symmetric off-diagonal exchange $\Gamma_1$ show that magnetic ground state is the [77$\bar{2}$] oriented zigzag AFM order. Further MC simulations indicate such zigzag AFM order is robust and not destroyed by the third NN AFM Kitaev interaction $K_3$.

The NN symmetric off-diagonal exchange and the NN FM Kitaev interaction cooperatively determine the preferred direction of magnetic moments of the zigzag AFM order. Obviously, the preferred direction is determined by the anisotropic Kitaev interaction and symmetric off-diagonal exchange. To make clear how it is picked out by them, we assume magnetic moments are parallel or antiparallel to the arbitrary unit vector $e_{xyz}=(x\ y\ z)$ in the $(x, y, z)$ coordinate. Then the dependence of the energy on the vector $e_{xyz}$ is

$$E = 2J_1 - 6J_3 + 2K_1\left(x^2 + y^2 - z^2\right) - 2K_3 - 4\Gamma_1\left(xy - yz - zx\right) \quad (5).$$

Eq. (5) clearly indicates the preferred direction is determined by the NN symmetric off-diagonal exchange $\Gamma_1$ and the NN FM Kitaev interaction $K_1$. Surprisingly, the third NN AFM Kitaev interaction $K_3$ makes no contribution to determining the preferred direction. If the symmetric off-diagonal exchange $\Gamma_1$ is only taken into account, the zigzag AFM order prefers being along the optimal [111] direction. Similarly, the zigzag order AFM order prefers lying in the $xy$ plane if the NN FM Kitaev interaction $K_1$ is only considered. If both are taken into consideration, the zigzag AFM order prefers being along the optimal [77$\bar{2}$] direction.

The experimentally measured dynamical response is captured by the $J_1$-$K_1$-$\Gamma_1$-$J_3$-$K_3$ model. Shown in FIG. 3a is the linear spin wave theory (LSWT) calculated spin wave spectrum. This spectrum has a minimum 2.22 meV in the vicinity of the **M**=(0.5 0.5 0.0) point, which is consistent with the fact that the experimentally measured low-energy magnetic mode $M_1$ has structured scattering near $Q$=0.62 Å$^{-1}$ at the lowest-energy 2.25 meV [2]. More importantly, the LSWT calculated powder-averaged scattering captures the experimentally observed magnetic modes $M_1$ and $M_2$ (FIG. 3b). The calculated energy positions of the low-energy magnetic mode $M_1$ and the high-energy magnetic mode $M_2$ locate at $E_1$=5.0 meV and $E_2$=6.6 meV, respectively, which are consistent with the experimentally measured $E_1$=4.1 meV and $E_2$=6.5 meV [2]. Additionally, the experimentally observed concave nature of the edge of the lower mode in (***Q***, E) space is reproduced (FIG. 3b). Further study shows if the NN symmetric off-diagonal exchange $\Gamma_1$ is slightly renormalized to 3.0 meV from 3.8 meV, the spin wave spectrum will have a minimum 2.41 meV near **M** point and the $E_1$ of the $M_1$ mode and the $E_2$ of the $M_2$ mode will be 4.3 meV and 6.3 meV, respectively. These are highly closed to the experimental observations. Hence the $J_1$-$K_1$-$\Gamma_1$-$J_3$-$K_3$ model nicely captures the dynamical response measurements of $\alpha$-RuCl$_3$.

The NN symmetric off-diagonal exchange is central to the presence of the spin wave gap. Here the [77$\bar{2}$] direction is approximated by the [110] one for simplicity and we consider two points in the reciprocal space, the **Γ**=(0.0 0.0 0.0) point and the **M**=(0.5 0.5 0.0) point. At the Γ point, LSWT predicts four branches of energy (Sec. IV of Supplemental Material):

$$E_1^\Gamma = \sqrt{\Gamma_1(\Gamma_1 + 2J_1 + 6J_3 + 2K_3)/2} \qquad (6.1),$$

$$E_2^\Gamma = \sqrt{(J_1 + 3J_3 + K_3)(\Gamma_1 - 2K_1)/2} \qquad (6.2),$$

$$E_3^\Gamma = \sqrt{(\Gamma_1 - 2J_1 + 6J_3 + 2K_3)(\Gamma_1 - 2J_1 - K_1)/2} \qquad (6.3),$$

$$E_4^\Gamma = \sqrt{(J_1 - 3J_3 - K_3 + K_1)(4J_1 - \Gamma_1 + 2K_1)/2} \qquad (6.4).$$

At the **M** point, LSWT predicts two doubly degenerate energies (Sec. IV of Supplemental Material) and the lowest one of them is

$$E_1^M = E_2^M = \frac{1}{2}\sqrt{-\Gamma_1^2 - \Gamma_1(J_1 - 3J_3 - K_3 + K_1) + K_1^2 + J_1 K_1 - 3J_3 K_1 - K_1 K_3 - \Delta} \qquad (7).$$

In Eq. (7), the parameter Δ is dependent on the Heisenberg, Kitaev interactions and off-diagonal exchange (Sec. IV of Supplemental Material). Eq. (6.1) and (7) show that energies $E_1^\Gamma$ and $E_1^M$ ($E_2^M$) are zero in the absence of the NN symmetric off-diagonal exchange $\Gamma_1$, indicating that the spin wave spectrum inevitably has no gap. Such non-gapped spin wave spectrum is consistent with Banerjee *et al*. theoretical results [2] since no symmetric off-diagonal exchange is included in their Hamiltonian. Therefore the NN symmetric off-diagonal exchange is necessary to the presence of spin wave gap.

*Discussion and summary.*—Here we discuss how the Kitaev QSL can be realized in *α*-RuCl₃ in the future experiments. If the NN Kitaev interactions $K_1$ and Heisenberg interaction $J_1$ are only taken into consideration, exact diagonalization study on a 24-site cluster shows the Kitaev QSL survives in a window from $J_1/|K_1|$ = -0.17 to $J_1/|K_1|$ = 0.12 around the exact Kitaev point (FIG. 4a) [40]. The ratio of $J_1/|K_1|$ in the $J_1$-$K_1$-$\Gamma_1$-$J_3$-$K_3$ model is -0.1698, indicating that *α*-RuCl₃ has the proximate Kitaev QSL ground state. Further study shows that such state persists even though the third NN Kitaev interaction $K_3$ is involved (FIG. 4b). However, the proximate Kitaev QSL breaks down if either the NN symmetric off-diagonal exchange $\Gamma_1$ or the third NN Heisenberg interaction $J_3$ is further considered (FIG. 4c and 4d). Especially, the proximate Kitaev QSL is extremely fragile against the NN symmetric off-diagonal exchange $\Gamma_1$. Our work indicates that it is crucial to weaken the NN symmetric off-diagonal exchange $\Gamma_1$ and the third NN Heisenberg interaction $J_3$ to realize the Kitaev QSL in $\alpha$-RuCl$_3$ in the future experiments. Such weakening may be achieved by applying strain, hydrostatic pressure or isovalent ions doping etc.

In summary, to understand the nature of the proximate Kitaev QSL in $\alpha$-RuCl$_3$, we propose a novel method to extract the magnetic interaction parameters in the spin-orbit coupled systems by constraining the direction of orbital moments. With this method, we have successfully unveiled the magnetic interactions of $\alpha$-RuCl$_3$ and thus propose a minimal $J_1$-$K_1$-$\Gamma_1$-$J_3$-$K_3$ model which well captures the experimental observations. Importantly, we propose the Kitaev QSL could be realized by suppressing the NN symmetric off-diagonal exchange and the third NN Heisenberg interaction. We demonstrate the present scheme is efficient for studying the magnetism of $J_{\text{eff}}$=1/2 magnets and give hints to future experiments for realizing the Kitaev QSL in $\alpha$-RuCl$_3$.

This paper was partially supported by the National Natural Science Foundation of China, the Special Funds for Major State Basic Research (2015CB921700), Program for Professor of Special Appointment (Eastern Scholar), Qing Nian Ba Jian Program, and Fok Ying Tung Education Foundation. We thank Y. L. Zhang for useful discussions.

**Figures**

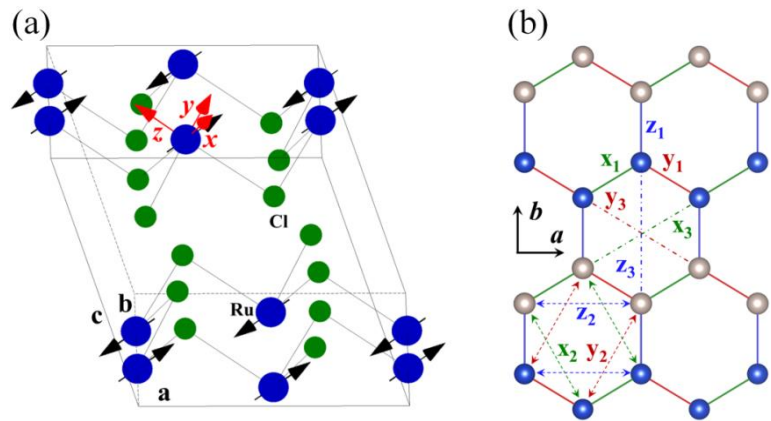

FIG. 1 (Color online) (a) Crystal structure and zigzag AFM order of $\alpha$-RuCl$_3$. Ru and Cl atoms are represented by the blue and green spheres, respectively. The $(x, y, z)$ coordinate expressing the HK Hamiltonian is drawn and its $x$-, $y$-, and $z$-axis point along three NN Ru–Cl bonds in a RuCl$_6$ octahedron. Black arrows represent magnetic moments. (b) Magnetic interaction paths. The NN, second NN and third NN x, y and z Ru-Ru paths are shown by green, red, and blue lines. Blue and white gray spheres have up and down spins, respectively.

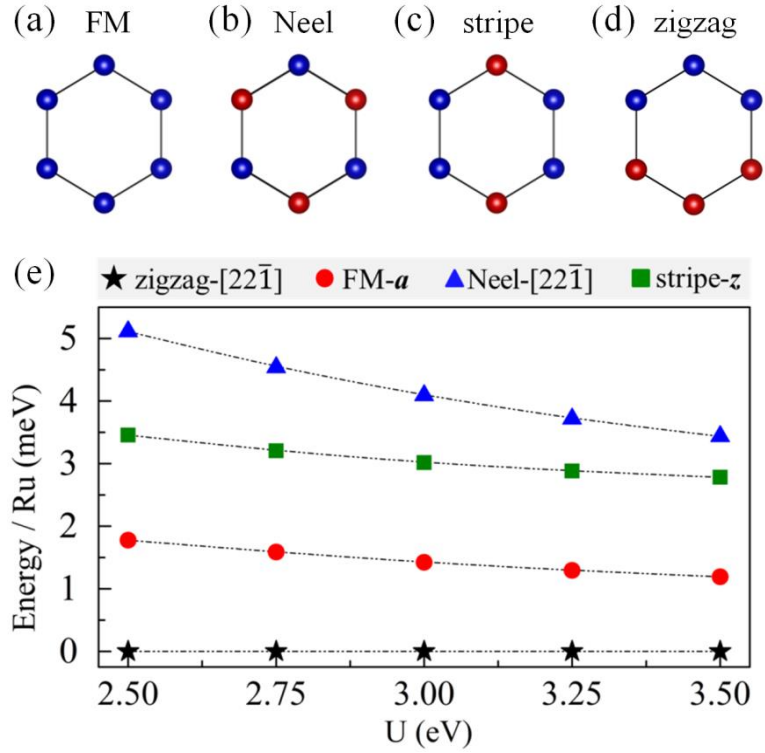

FIG. 2 (Color online) Illustration of different magnetic orders (a) FM, (b) Neel AFM and (c) stripe AFM and (d) zigzag AFM. In (a), (b), (c) and (d), blue and red spheres have up and down spins, respectively. (e) Energy differences per Ru atom of the considered magentic orders are plotted with respect to U. The zigzag-$[22\bar{1}]$ (black star) AFM has the lowest energy and are set as the energy reference. As for the FM, Neel AFM and stripe AFM, only their lowest energies are plotted.

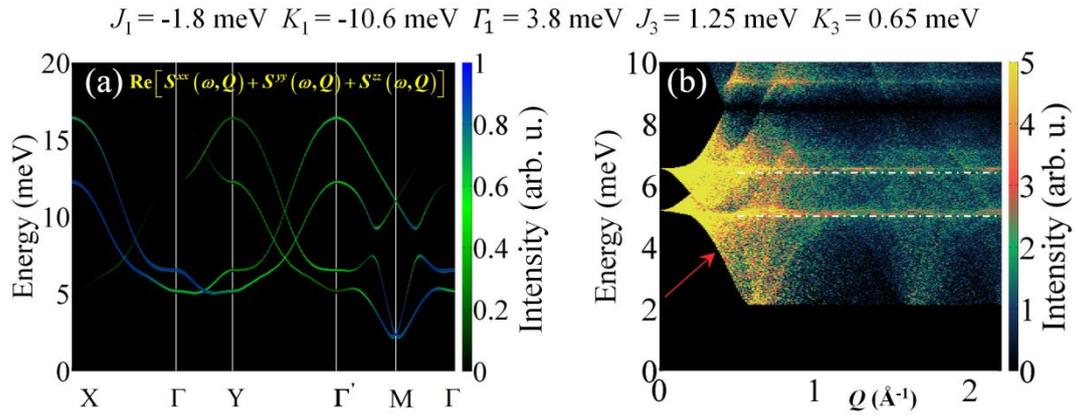

FIG. 3 (Color online) (a) Spin wave spectrum and (b) powder-averaged scattering of the $J_1$-$K_1$-$\Gamma_1$-$J_3$-$K_3$ model calculated with the LSWT. The adopted magnetic interaction parameters are listed in the legend. In (b), the red arrow highlights the experimentally observed concave nature of the edge of the lower mode in ($\mathbf{Q}$, E) space [2] and the white dash-dotted lines show the energy positions of the magnetic modes $M_1$ and $M_2$.

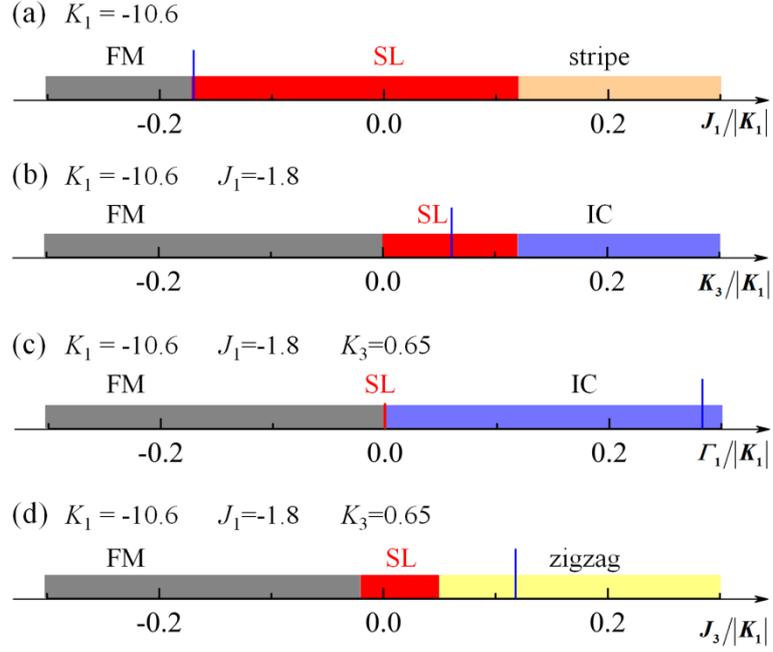

FIG. 4 (Color online) Exact diagonalization study of the breakdown of the Kitaev QSL due to (a) the NN Heisenberg interaction $J_1$, (b) the third NN Kitaev coupling $K_3$, (c) the NN symmetric off-diagonal exchange $\Gamma_1$ (3.0 meV is adopted here) and (d) the third NN Heisenberg interaction $J_3$. SL and IC denote "spin liquid" and "incommensurate", respectively. The magnetic parameters shown in the legend are in unit of meV. The blue vertical lines in (a)-(d) show the ratio $J_1/|K_1|$, $K_3/|K_1|$, $\Gamma_1/|K_1|$ and $J_3/|K_1|$, respectively, which are calculated based on the magnetic interactions parameters listed in the last row in the Table I.

Table *I*. Calculated magnetic interaction parameters in unit of meV in the case of different U. Detailed magnetic interaction parameters are given in the Sec. III of Supplemental Material. The letters x, y and z in the parentheses are the indicator of Ru-Ru bonds. The last row is the magnetic interaction parameters of the $J_1$-$K_1$-$\Gamma_1$-$J_3$-$K_3$ model.

| U (eV) | $J_1$ (z) | $J_1$ (x/y) | $K_1$ (z) | $K_1$ (x/y) | $\Gamma_1^z$ (z) | $\Gamma_1^{x/y}$ (x/y) | $J_3$ (z) | $J_3$ (x/y) | $K_3$ (z) | $K_3$ (x/y) |
|---|---|---|---|---|---|---|---|---|---|---|
| 2.5 | -2.1 | -2.5 | -13.9 | -14.7 | 6.5 | 6.4 | 2.0 | 2.1 | 0.9 | 0.9 |
| 3.0 | -1.8 | -2.2 | -11.9 | -12.4 | 4.9 | 4.8 | 1.6 | 1.6 | 0.8 | 0.7 |
| 3.5 | -1.6 | -2.0 | -10.4 | -10.8 | 3.8 | 3.8 | 1.2 | 1.3 | 0.7 | 0.6 |
| Model | $J_1$ -1.8 | | $K_1$ -10.6 | | $\Gamma_1$ 3.8 | | $J_3$ 1.25 | | $K_3$ 0.65 | |